\documentclass[letterpaper,11pt]{article}
\pdfoutput=1
\usepackage{jheppub_mod}
\usepackage{graphicx, amsmath, amssymb, amstext,xcolor,txfonts}
\usepackage{relsize, float, multirow}
\usepackage{array, epstopdf, hyperref}
\usepackage{caption, subcaption, mathtools}
\usepackage{newtxmath,ulem}
\usepackage{multicol}

\definecolor{seagreen}{rgb}{0.18, 0.55, 0.34}

\title{Probing inflationary features with galaxy ultraviolet luminosity function observables }

\author[a,1]{Sandeep Kumar Acharya\note{Corresponding author.}}

\affiliation[a]{Indian Institute of Astrophysics, Koramangala II Block, Bengaluru 560 034, India}

\emailAdd{sandeepkumaracharya92@gmail.com}
\date{\today}

\abstract{We use the galaxy ultraviolet luminosity function measurements at $z=4-8$ to constrain modification to standard inflationary power spectrum. These observables are sensitive to the matter power spectrum which itself depends on inflationary initial conditions. We consider a specific model where a bump feature is introduced to the standard power law inflation spectrum. We find that the galaxy luminosity observables can probe such modifications at wavenumbers $0.5\lesssim k \lesssim 50$ Mpc$^{-1}$. We obtain upper limits on the amplitude of bump-like features at the mentioned wavenumbers. We obtain constraints which are similar to previous constraints on these models using measurements of optical depth of reionization. However, the galaxy luminosity functions are a more direct probe for these type of models and, therefore, can complement indirect constraints coming from measurements of IGM properties.   }

\notoc
\begin{document}
\maketitle
\newpage

\section{Introduction}
The understanding of properties and distribution of galaxies at high redshifts, especially during reionization epoch ($z\gtrsim 6$), has become a subject of significant interest in recent years. With the surge of observational data from large galaxy surveys, we are beginning to test our theoretical understanding of high redshift galaxy formation, developed over last couple of decades  \citep{BL2001,Millenium2009,MFC2011,Illustris2014,EAGLE2015,DF2018,Universemachine2020,MC2022,MP2026}. While measurements of Cosmic Microwave Background (CMB) \citep{WMAP2011,Planck2020,ACT2025,SPT2025} and large scale structure (LSS) \citep{eBOSS2021,DESI2025} have established our understanding of the Universe on larger length scales ($\gtrsim 1-10$ Mpc), these high redshift galaxies provide a complementary probe on smaller length scales. 

One such high redshift observable is galaxy ultraviolet luminosity function (UVLF) which is defined as the comoving number density of galaxy per unit luminosity. These measurements are made at rest frame wavelength of about 1500\AA. The observed luminosity provides a proxy for the inference of star formation rate which captures information regarding galaxy formation and their cosmological evolution. The Hubble Space Telescope (HST) has measured UVLFs upto $z\approx 10$ \citep{Bouwens2015,Atek2018,Bouwens2021}. Recently, James Webb Space Telescope (JWST) has pushed this limit to $z\approx 15$ \citep{Donnan2023,Harikane2023,Bouwens2023,McLoed2024}. These measurements have inferred a large abundance of UV bright galaxies as compared to the prediction from standard galaxy formation model in  $\Lambda$CDM framework extrapolated to these redshifts \citep{BCWHWMYSGYE202,Vogelsberger2020,Kannan2023,LHHTW2023,MF2023,MTT2023}. Several proposals beyond $\Lambda$CDM cosmology, higher star formation efficiency as well as stochasticity of bright UV galaxies have been proposed \citep{B2023,Padmanabhan2023,PL2023,DSBML2023,MV2023,LB2022,BFR2023,HRUV2023,GYCC2023,SVBTK2023,SL2024} to alleviate this problem. More recent works \cite{CC2024,DFGBTPDSB2025} show that JWST data necessitates evolving star formation efficiency as a function of redshift within $\Lambda$CDM framework.

The UVLF observables have also been used to probe particle nature of dark matter  \citep{DG2024,LKBM2025,DMSL2025,BCLMR2025,BLBMC2025}. Non-trivial particle interaction of dark matter beyond cold dark matter paradigm modifies the shape of matter power spectrum. This modification imprints its characteristic signature on halo mass function which captures the comoving number density of dark matter halo as a function of halo mass. Since the dark matter halos act as host for galaxies, the number density of galaxies including total star formation rate in the universe is also modified. Therefore, UVLF measurements which constrain star formation can indirectly test interaction properties of dark matter. Further, the matter power spectrum also captures information regarding initial condition or inflationary power spectrum \citep{Guth1981,Linde1982,Straobinsky1980,AS1982} of the Universe. Therefore, these observables can be used to probe modification to our understanding of inflation even within standard $\Lambda$CDM framework. Current CMB and LSS measurements are consistent with inflation power spectrum being a featureless power law within wavenumber $10^{-4}$ Mpc$^{-1}\lesssim k \lesssim 0.1$ Mpc$^{-1}$ \citep{WMAP2011,Planck2020,ACT2025,SPT2025,eBOSS2021,DESI2025}. However, several models predict scale dependent features \citep{BH2009,JCGSS2009,JSSS2009,HAJSS2010,MHA2012,CHP2015} which can modify the power spectrum at $k\gtrsim 0.1$ Mpc$^{-1}$. The authors in \cite{YOTT2020} have used the UVLF data to study modification to standard inflationary power spectrum with some choice of model parameterization such as adding running parameters or step-like features. 

In this paper, we probe the effect of modified inflationary power spectrum on the UVLF observables for a specific model which introduces a bump-like feature on top of power law spectrum due to particle prediction during inflation \citep{CKRT2000,BHKP2009,Furuuchi2016,PPS2017,FNJ2020}. These models have been studied in the context of CMB observables \citep{NFC2022} and galaxy two point correlation function \citep{BFMMV2023} and upper limits on the amplitude of such features have been obtained within $10^{-4}$ Mpc$^{-1}\lesssim k \lesssim 0.1$ Mpc$^{-1}$ regime. Recently, this model was studied in the context of predictions for 21cm observables \citep{NCSMF2025,VNC2026}. It was shown that global 21cm signal can constrain these features at wavenumbers $0.1$ Mpc$^{-1}\lesssim k \lesssim 10$ Mpc$^{-1}$. In this work, we show that available UVLF data can already constrain the amplitude of bump like features at these larger wavenumbers.  

This paper is organized as follows. We describe our modelling of UVLF data in Sec. \ref{sec:UVLF_modelling}. Our formalism for modified inflationary power spectrum is described in Sec. \ref{sec:inflation}. We provide the methodology used and our results on these non-standard models in Sec. \ref{sec:data} and \ref{sec:results} respectively.  We conclude in Sec. \ref{Sec:conclusions}. Throughout the work, we use the best fit cosmological parameters obtained from {\it Planck}\citep{Planck2020}.

\section{Theoretical modelling of UVLF}
\label{sec:UVLF_modelling}
In order to theoretically model star formation inside galaxies, we use the formalism of halo model \citep{Seljak2000,PS2000}. In this formalism, the dark matter halos provide gravitational potential well for baryons to fall into after decoupling from CMB photons. Therefore, the baryonic tracers such as distribution of galaxies are expected to be somewhat of a proxy of dark matter distribution as well as its particle properties. 
In order to relate the stellar mass of galaxy ($M_*$) to the mass of dark matter halo ($M_h$), we employ the astrophysical model used in  \cite{SMB2022}. According to this model, the relation between $M_*$ and $M_h$ can be parameterized as,
\begin{equation}
    M_*=f_*(M_h,z)\left(\frac{\Omega_b}{\Omega_m}\right)M_h,
    \label{eq:stellar_mass}
\end{equation}
where $f_*$ is the star formation efficiency which is parameterized as,
\begin{equation}
f_*(M_h,z)=\frac{\tilde{f}_*}{\left(\frac{M_h}{M_c}\right)^{-\alpha_*}+\left(\frac{M_h}{M_c}\right)^{-\beta_*}}
\end{equation}
Further, the parameters ${\tilde f}_*, M_c$ evolve with redshifts as $\tilde{f}_*=\tilde{f}_{*,12}\left[\frac{1+z}{7}\right]^{\epsilon_{\alpha}}$ and $M_c=M_{c,12}\left[\frac{1+z}{7}\right]^{\epsilon_{\beta}}M_{\odot}$ where $\tilde{f}_{*,12}$ and $M_{c,12}$ are respective normalization at $M_h=10^{12}$ $M_{\odot}$. Therefore, our fiducial astrophysical model has 6 parameters $\tilde{f}_{*,12},M_{c,12},\alpha_*,\beta_*,\epsilon_{\alpha}, \epsilon_{\beta}$. The star formation rate (SFR) is given by,
\begin{equation}
    \dot{M_*}(M_h,z)=\frac{M_*(M_h,z)}{t_*(z)},
\end{equation}
where $t_*(z)$ is the timescale of star formation which we assume to be Hubble scale or $t_*(z)=1/H(z)$. There can be O(1) numerical factor in the above relation which can be absorbed within $\tilde{f}_{*}$. Therefore, the expression for SFR becomes,
\begin{equation}
    \dot{M_*}(M_h,z)=f_*(M_h,z)H(z)\left(\frac{\Omega_b}{\Omega_m}\right)M_h
\end{equation}
The luminosity of galaxy at rest frame wavelength $1500\AA$  can be related to its SFR via the scaling relation, 
\begin{equation}
{\mathcal{K}}_{UV}=\frac{\dot{M_*}(M_h,z)}{L_{UV}},
\end{equation}
which is tuned to observations at low redshifts and with the assumption of Salpeter IMF \citep{MD2014}. We assume a fiducial value of $\mathcal{K}_{UV}=1.15485\times 10^{ -28} M_{\odot}$ ${\rm yr}^{-1}$/ ergs s$^{-1}$ Hz$^{-1}$ \citep{MD2014}. Therefore, the UV luminosity can be expressed as,
\begin{equation}
L_{UV}=\frac{f_{*}(M_h,z)}{{\mathcal{K}}_{UV}}H(z)\left(\frac{\Omega_b}{\Omega_m}\right)M_h
\end{equation}
We can relate the luminosity to the UV magnitude using the relation,
\begin{equation}
{\rm log_{10}}\left(\frac{L_{UV}}{{\rm ergs}\hspace{0.1cm} {\rm s}^{-1}\hspace{0.1cm}{\rm Hz}^{-1}}\right)=0.4\times (51.6-M_{UV})
\end{equation}
Finally, the UVLF or the number density of galaxies as a function of their luminosity or magnitude is given by,
\begin{equation}
\Phi_{UV}(z,M_{UV})=\frac{{\rm d}n}{{\rm d}M_{UV}}=\frac{{\rm d}n}{{\rm d}M_{h}}\frac{{\rm d}M_h}{{\rm d}L_{UV}}\frac{{\rm d}L_{UV}}{{\rm d}M_{UV}}
\label{Eq:Phi_theory}
\end{equation}
The above eq. provides an average relation between UVLFs and the magnitude. In principle, there is some scatter or stochasticity around this average relation. The authors in \cite{SMB2022} parametrized this scatter as a gaussian function around the average value. It has been shown that this scatter can have important consequences for JWST data 
\citep{MCSSMBVQSE2026,SFHSWC2023}. However, for HST data (considered in this work), the best fit standard deviation of this gaussian function turns out to be much smaller than 1 \citep{SMB2022}. Therefore, the effect of scatter may not be important for HST data. In this work, we have not taken this factor into account.

The halo mass function $\frac{{\rm d}n}{{\rm d}M_{h}}$ encodes the comoving number density of dark matter halos in the mass range $M_h$ and $M_h+{\rm d}M_h$. It is a function of variance of matter density field $\sigma(M_h,z)$ and is given by \citep{PS1974},
\begin{equation}
    \frac{{\rm d}n}{{\rm dln}M_h}=\frac{\rho_m}{M_h}\frac{{\rm dln}\sigma^{-1}}{{\rm dln}M_h}f(\sigma,z),
\end{equation}
where $\rho_m$ is the average matter energy density today. In this work, we use the fitting function $f(\sigma,z)$ provided by Jenkins et al \citep{Jenkins2001}. In literature, there are different choices of fitting functions which have been used. These different choices add some uncertainty to the analysis. We have also done some calculations  using the fitting function by Tinker et al \citep{Tinker2008} for comparison. One specific case of parameter estimation is shown in Fig. \ref{fig:corner_k_1.0_comp}. We have checked that our constraints on bump-like inflationary features do not change significantly by changing different fitting functions for dark matter halo distribution.

The variance of matter field at $z$ can be obtained as, $\sigma(M_h,z)=\sigma(M_h)D(z)$, where $D(z)$ is the growth factor which depends upon the assumed cosmological model. We have used the {\it HMF} code \citep{HMF2013} to compute the halo mass function as well as {\it Planck} \cite{Planck2020} best fit cosmological parameters to compute the growth factor. The smoothed variance $\sigma(M_h)$ at $z=0$ is expressed as,
\begin{equation}
    \sigma(R)=\frac{1}{2\pi^2}\int_0^{\infty}{\rm d}k k^2P_m(k)W^2(kR),
    \label{eq:sigma_R}
\end{equation}
where $P_m(k)$ is the matter power spectrum at $z=0$ , $W(kR)$ is top-hat smoothing function in $R$ or position space and $M_h=\frac{4\pi}{3}\rho_mR^3$.

\section{Modification to matter power spectrum due to inflationary bump like feature}
\label{sec:inflation}
\begin{figure}
\centering
\includegraphics[scale=0.4]{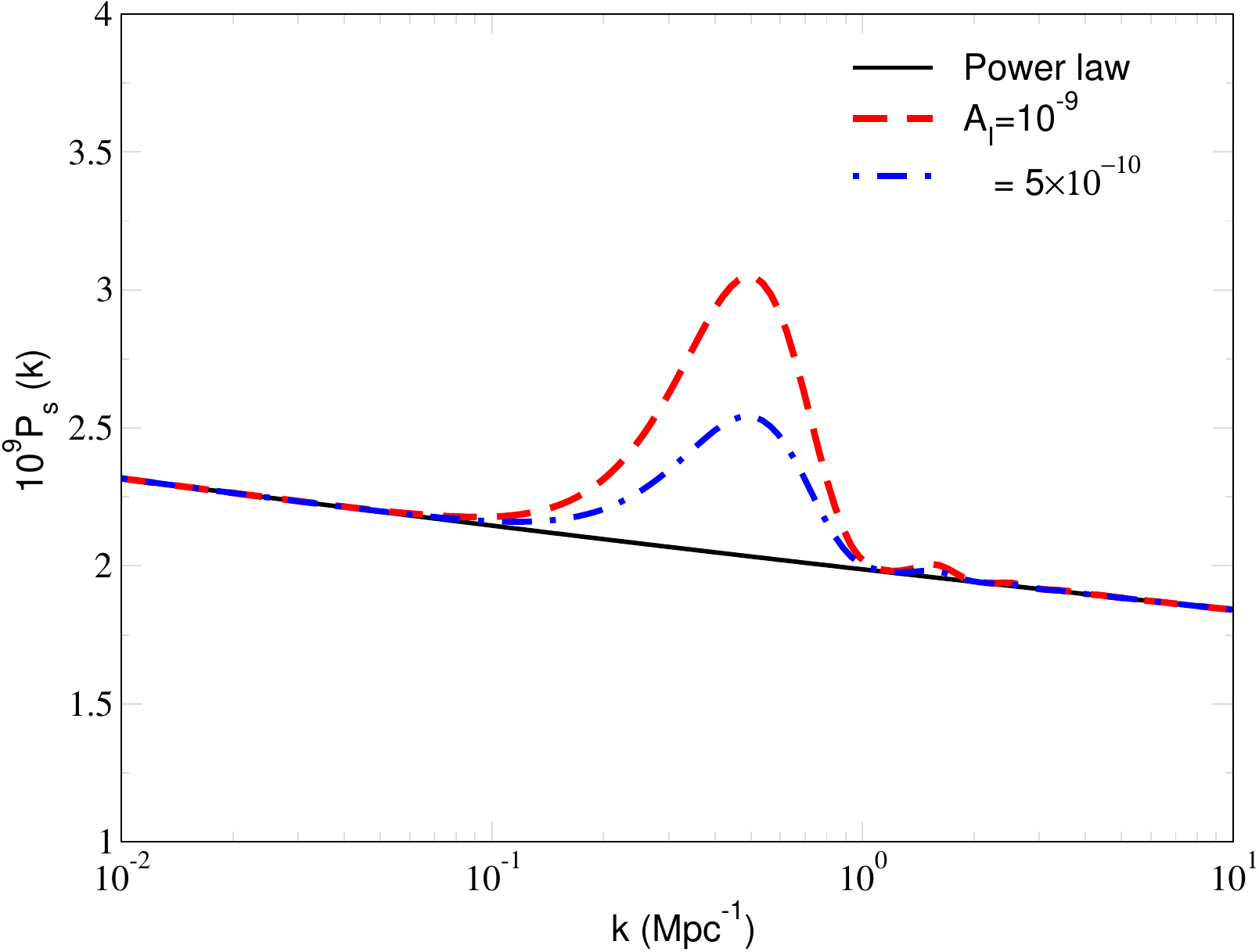}
\caption{Modified inflationary power spectrum with a single bump like feature, $k_{\rm peak}=0.5$ Mpc$^{-1}$ and amplitude as denoted in the plot. We plot the standard power law spectrum in black with parameters $A_s=2.1\times 10^{-9}$ and $n_s=0.9649$. }
\label{fig:Ps}
\end{figure}

\begin{figure}
\centering
\includegraphics[scale=0.4]{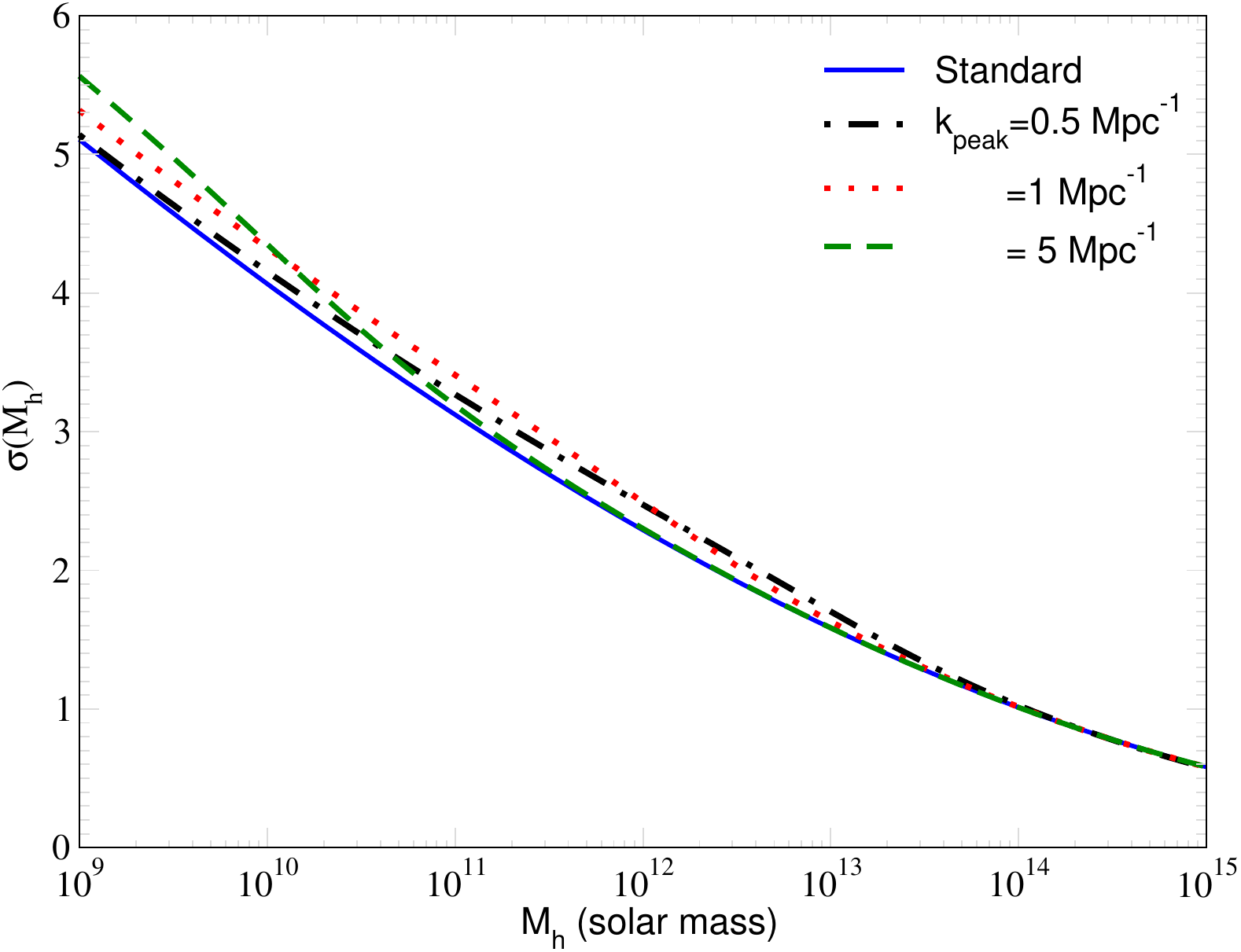}
\caption{$\sigma(M_h)$ as a function of halo mass at $z=0$ for different $k_{\rm peak}$ and $A_{I,-9}=1$. We have fixed the parameters $A_s=2.1\times 10^{-9}$ and $n_s=0.9649$. }
\label{fig:sigma_Mh}
\end{figure}

The standard inflationary power spectrum is parameterized as,
\begin{equation}
    P_{\rm std}(k)=A_s\left(\frac{k}{k_*}\right)^{n_s-1},
\end{equation}
with $k_*=0.05$ Mpc$^{-1}$, $A_s$ and $n_s$ parameterizes the amplitude and slope of power law. They are tightly constrained by CMB data \citep{Planck2020} with values ${\rm ln}(10^{10}A_s)=3.044\pm 0.014$, $n_s=0.9649\pm 0.0042$. Bursts of particle production during inflation can generate bump-like features in the power spectrum \citep{CKRT2000,BHKP2009,Furuuchi2016,PPS2017,FNJ2020}. We parameterize the modified power spectrum by including the dominant term as \citep{PPS2017,NCSMF2025},
\begin{equation}
P_s(k)=A_s\left(\frac{k}{k_*}\right)^{n_s-1}+A_I\sum_i\left(\frac{f_1(x_i)}{f_1^{\rm max}}\right)
\label{eq:Ps_eqn}
\end{equation}
The function $f_1(x)$ is expressed as,
\begin{equation}
    f_1(x_i)=\frac{[{\rm sin}(x_i)-{\rm SinIntegral}(x_i)]^2}{x_i^3},
\end{equation}
where $x_i=\frac{k}{k_i}$ and $f_1^{\rm max}$ is the maximum value of $f_1(x)$. The parameter $k_i$ is related to the position of $i$th feature which peaks at $k_{{\rm peak},i}=3.35k_i$. For simplicity, we consider one bump or $i=1$ in this work. In Fig. \ref{fig:Ps}, we plot the modified power spectrum with $k_{\rm peak}=0.5$ Mpc$^{-1}$ for a couple of cases. We compute the modified matter power spectrum as $P_m(k)=P_s(k)\mathcal{T}^2(k)$, where $\mathcal{T}(k)$ is the matter transfer function computed using {\it CLASS} \citep{BLT2011}. We substitute the matter power spectrum in Eq. \ref{eq:sigma_R} in order to compute $\sigma(M_h)$. In Fig. \ref{fig:sigma_Mh}, we plot $\sigma(M_h)$  for few cases with varying $k_{\rm peak}$ and fixed amplitude $A_I$. We find that $\sigma(M_h)$ changes non-negligibly which will be further magnified in the halo mass function due to its exponential dependence on $\sigma(M_h)$ \citep{Jenkins2001}. In the rest of the paper, we have introduced the parametrization $A_{I,-9}=\frac{A_I}{10^{-9}}$ in order to simplify the choice of range during the parameter estimation stage. 

\section{Data and analysis}
\label{sec:data}

\begin{table}[!htb]

      \centering
       \begin{tabular}{l|c|r} 
 parameters & prior & posterior values  \\
    \hline
    \vspace{0.1cm}
   $\tilde{f}_{*,12}$ & [0.05,1.0] & $0.571^{0.030}_{-0.034}$ \\ 
   \vspace{0.1cm}
   $M_{c,12}$ & [0.1,10] & $0.351^{0.206}_{-0.138} $ \\
   \vspace{0.1cm}
   $\alpha_*$ & [0,2] & $0.632^{0.151}_{-0.119}$ \\
   \vspace{0.1cm}
   $\beta_*$ & [-2,2] & $-0.557^{0.146}_{-0.170}$ \\
   \vspace{0.1cm}
   $\epsilon_{\alpha}$ & [-5,5] & $0.566^{0.205}_{-0.206}$ \\ \vspace{0.1cm}
   $\epsilon_{\beta}$ & [-5,5] & $-0.449^{0.649}_{-0.616}$ \\
   \hline
   $A_{I,-9}$ & [0.1,1000] &  \textminus
 \end{tabular}
 \caption{Prior range for parameters in our model and posterior values (95\% confidence) for the astrophysical parameters in our fiducial model with $A_I=0$ corresponding to Fig. \ref{fig:std_6params}.  }
 \label{tab:parameter_table}
    \end{table}

\begin{figure}
\centering
\includegraphics[scale=0.4]{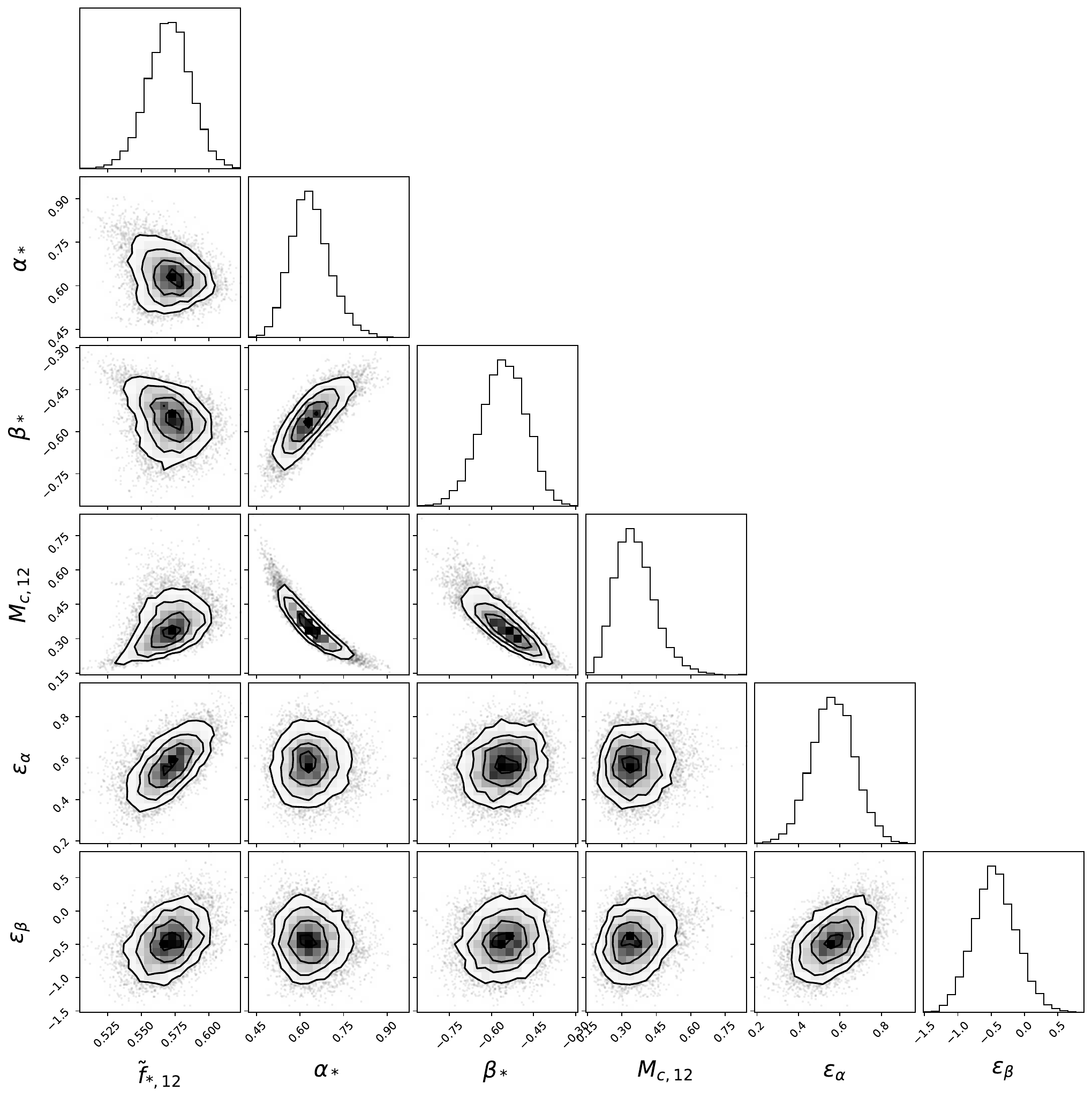}
\caption{Corner plot for our fiducial astrophysical model with $A_I=0$. }
\label{fig:std_6params}
\end{figure}

\begin{figure}
\centering
\includegraphics[scale=0.8]{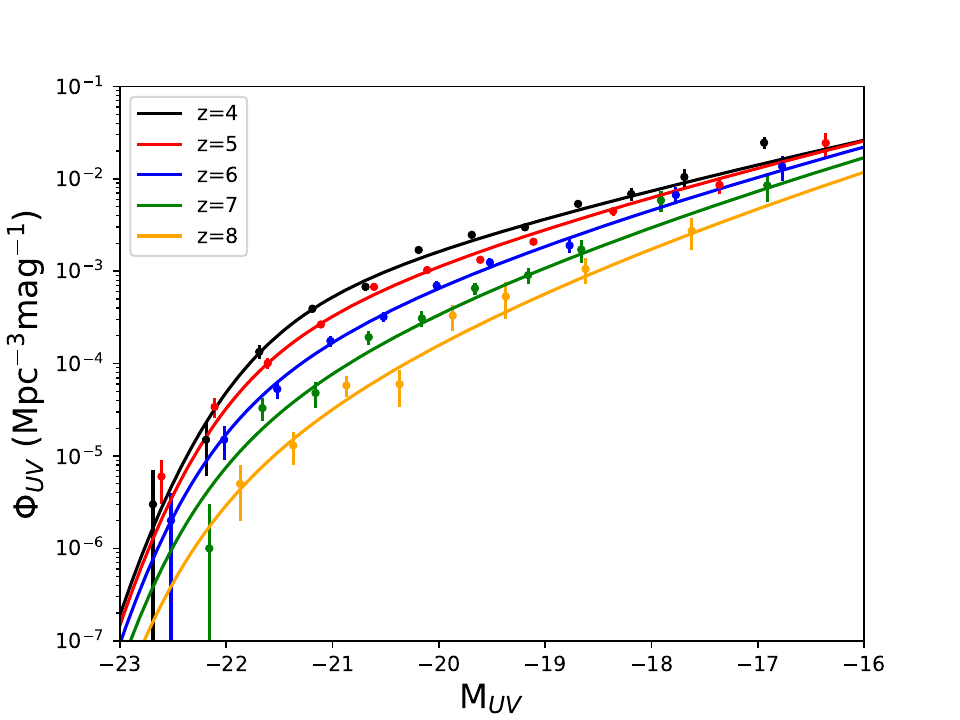}
\caption{Theoretical prediction for $\Phi_{UV}$ obtained using Eq. \ref{Eq:Phi_theory} using the median posterior values of our fiducial astrophysical model in Table \ref{tab:parameter_table}, along with the data used in this work. }
\label{fig:fit_fiducial}
\end{figure}

\begin{figure}
\centering
\includegraphics[scale=0.6]{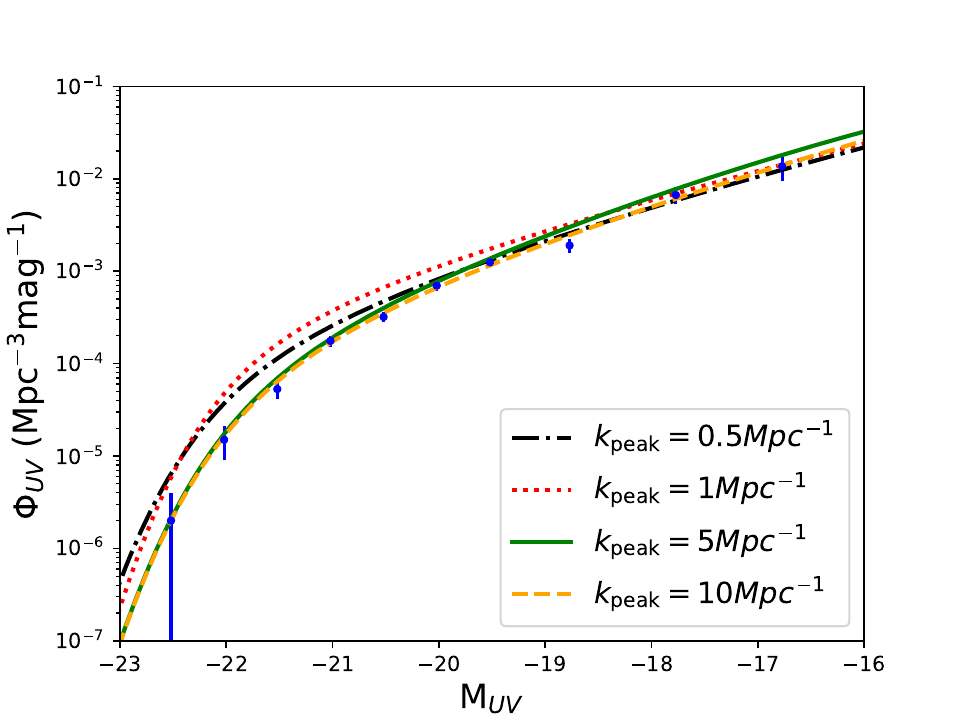}
\caption{Theoretical prediction for UVLF at $z=6$ for different $k_{\rm peak}$ and $A_{I,-9}=1$ using the median posterior values of  astrophysical parameters obtained in Table \ref{tab:parameter_table}.   }
\label{fig:AI_1e9}
\end{figure}

\begin{figure}
\centering
\includegraphics[scale=0.35]{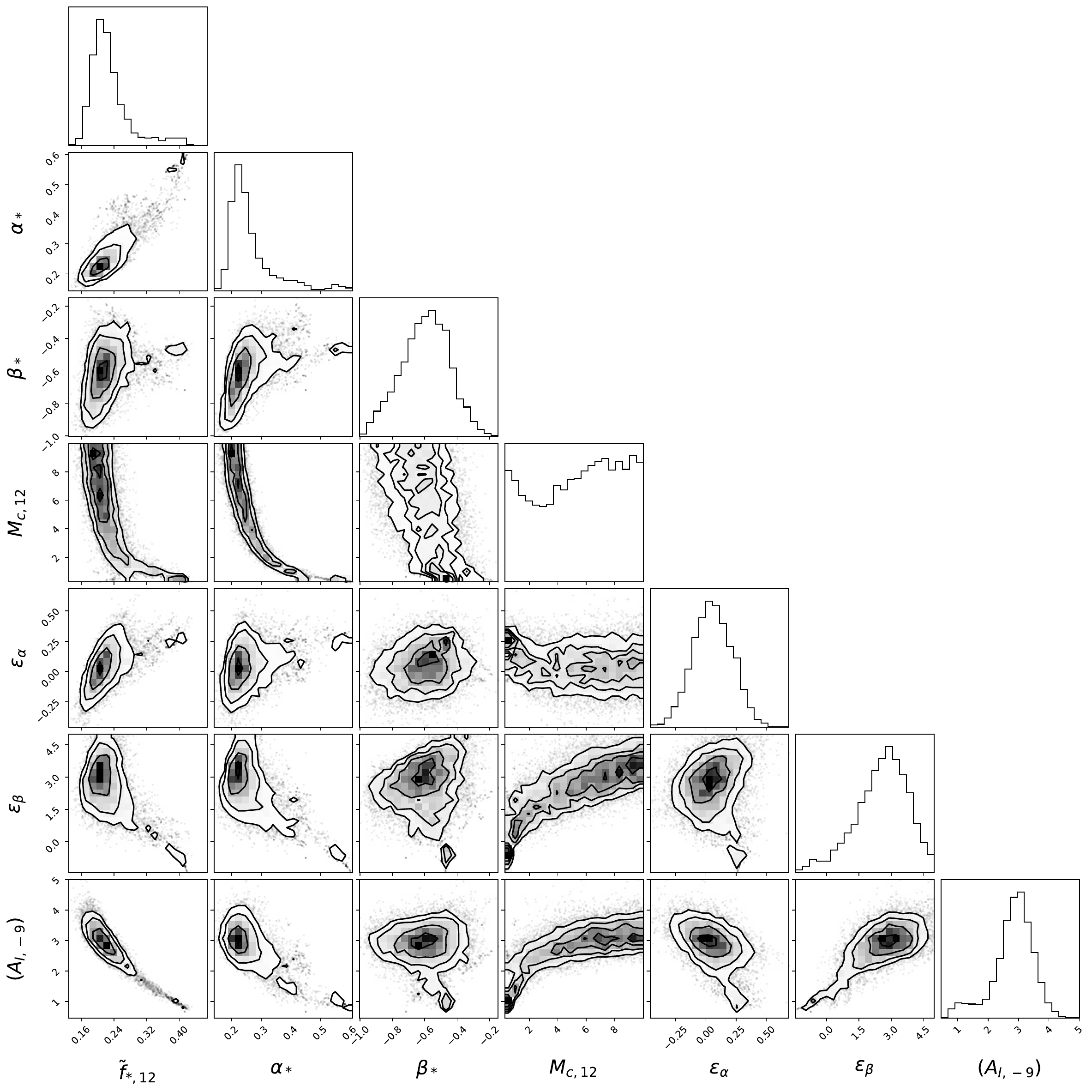}
\caption{Corner plot for our model with bump like feature at $k_{\rm peak}=1$ ${\rm Mpc^{-1}}$. }
\label{fig:7params_k_1.0}
\end{figure}

\begin{figure}
\centering
\includegraphics[scale=0.8]{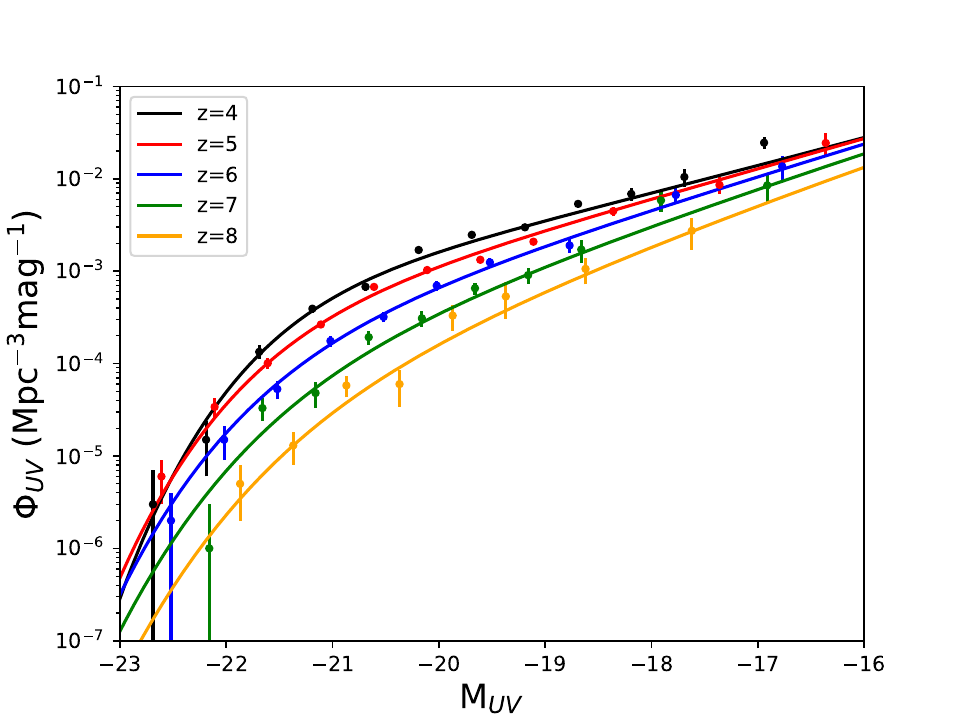}
\caption{Fit to the data for our model with non-zero $A_I$ and $k_{\rm peak}=1.0$ ${\rm Mpc^{-1}}$. The parameters are chosen to be the median values of posteriors shown in Fig. \ref{fig:7params_k_1.0}.   }
\label{fig:fit_k_1.0_AI}
\end{figure}

\begin{figure}
\centering
\includegraphics[scale=0.8]{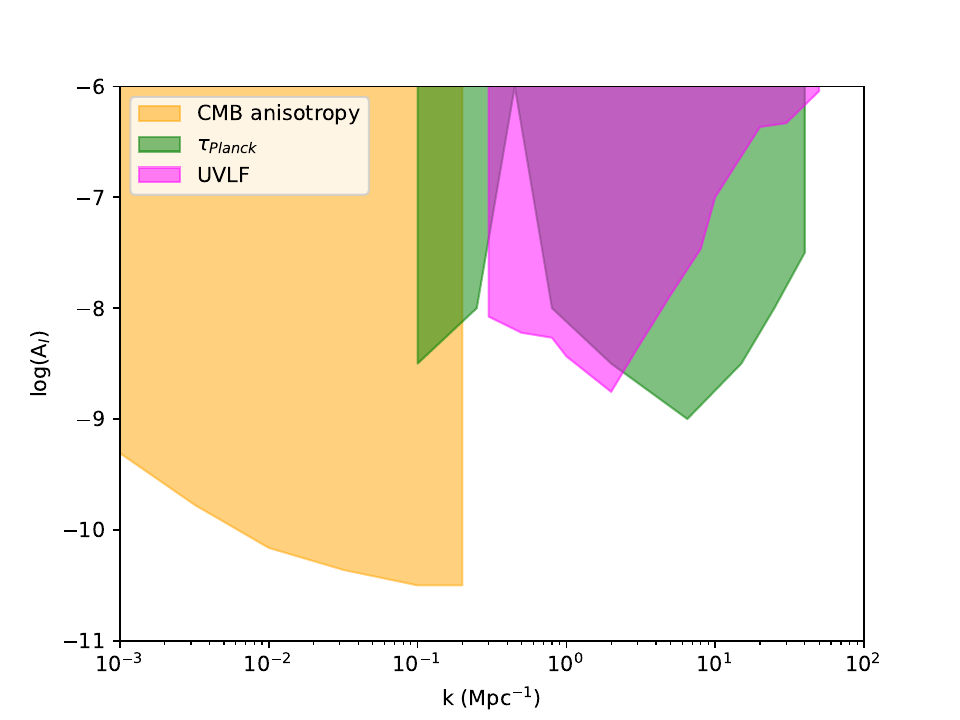}
\caption{Constraints on $A_I$ (at 95 percent confidence interval) as a function of scale from different probes as denoted in the plot. Our new constraints cover range between $k\in [0.3-50]$ Mpc$^{-1}$. Our constraints are competitive with constraints derived from CMB optical depth $\tau_{\rm Planck}$ \citep{NCSMF2025}. In deriving the optical depth constraint, a fiducial astrophysical model with fixed parameter values was assumed. }
\label{fig:UVLF_constraint}
\end{figure}

We use the UVLF measurements made with Hubble Space Telescope (HST) within the redshift range 4 to 8 to do our analysis. The measurements are obtained from Table 5 of \cite{Bouwens2015}. For completeness, we compile the data points used in this work in Table \ref{tab:table1}. We have excluded few data points with upper limits, for the analysis. At higher redshifts, measurements from JWST are available \citep{Donnan2023,Harikane2023,Bouwens2023,McLoed2024}. Recent analysis \citep{CC2024,DFGBTPDSB2025} show that JWST data require significant redshift evolution of star formation efficiency $f_*$ which may require more complex astrophysical modelling. Also, the data points becomes increasingly more sparse with higher uncertainty as we go to higher redshifts. We have not used JWST data in this analysis.

We have used Gaussian likelihood for parameter estimation. The log-likelihood is given as,
\begin{equation}
    -2{\rm ln}\mathcal{L}=\sum_{M_{UV},z}\left(\frac{\Phi_{UV}^{\rm data}-\Phi_{UV}^{\rm theory}}{\sigma_{\Phi_{UV}}}\right)^2
    \label{eq:log_likelihood}
\end{equation}
where $\Phi_{UV}^{\rm theory}$ is obtained from Eq. \ref{Eq:Phi_theory} and $\Phi_{UV}^{\rm data}$, $\sigma_{\Phi_{UV}}$ are given in Table \ref{tab:table1}. We sum over contribution from individual data points over $M_{UV}$ and redshifts bins. We use the publicly available {\it emcee} sampler \citep{emcee2013} to run Markov Chain Monte Carlo (MCMC) using the Gaussian likelihood described above. Our prior choices are shown in Table \ref{tab:parameter_table}. We choose flat prior on all the parameters within the denoted range.

\section{Results}
\label{sec:results}
As a first exercise, we fit our fiducial astrophysical model to the data. We have fixed the parameters $A_s=2.1\times 10^{-9}$ and $n_s=0.9649$ for the analysis. The posteriors and the corresponding best fit values are shown in Fig. \ref{fig:std_6params} and Table \ref{tab:parameter_table} respectively. We compare theoretical computation for $\Phi_{UV}$ using the best fit values of our model (Eq. \ref{Eq:Phi_theory})  to the data in Fig. \ref{fig:fit_fiducial}. In Fig. \ref{fig:AI_1e9}, we show theoretical prediction for $\Phi_{UV}$ at $z=6$ with an inflationary bump-like feature for different $k_{\rm peak}$ but a fixed amplitude and using the best fit astrophysical parameters. We see that these scenarios are already in tension with data assuming a fixed set of astrophysical parameters. 

Then, we proceed to jointly fit our non-standard parameter $A_{I,-9}$ along with the astrophysical parameters for different $k_{\rm peak}$. We show the posterior for our fitted model for $k_{\rm peak}=1$ Mpc$^{-1}$ in Fig. \ref{fig:7params_k_1.0} as an example. Interestingly, we find that the data prefers non-zero $A_I$ for $k_{\rm peak}=1$ ${\rm Mpc^{-1}}$. However, for other value of $k_{\rm peak}$, we find that the data is consistent with $A_I=0$. We show the posteriors for a couple of cases with $k_{\rm peak}=0.5$ and 5 ${\rm Mpc^{-1}}$ in Fig. \ref{fig:corner_k_0.5_10}. Non-zero $A_I$, with fixed astrophysical parameters, increases matter variance (Fig. \ref{fig:sigma_Mh})  boosting $\Phi_{UV}$ through the dark matter halo mass function (Eq. \ref{Eq:Phi_theory}) which brings theoretical prediction in tension with data.  

We plot the constraints or allowed upper limits on $A_I$ using UVLF data in Fig. \ref{fig:UVLF_constraint}. Our constraints cover range $k\in [0.3-50]$ Mpc$^{-1}$. At larger length scales or smaller $k$, CMB anisotropy provides the strongest constraint \citep{NFC2022}. The authors used {\it Planck2018} \citep{Planck2020} temperature and polarization data to obtain these constraints using {\it Plik-lite} likelihood. Our constraints are competitive with those derived from CMB optical depth or $\tau_{\rm Planck}$ \citep{NCSMF2025} in the relevant range of $k$. We note that to obtain the $\tau$ constraint, a fiducial astrophysical model with fixed parameter values were assumed. Further, $\tau$ is related to the ionization history of the Universe which depends on the Intergalactic medium (IGM) properties. Ionizing photons from stars inside the smallest galaxies, which are too faint to see, escape to IGM and affect the ionization history of the Universe. Therefore, the $\tau$ constraint is an indirect probe of galaxy properties and capture the cumulative effect of smallest to largest galaxies. However, UVLFs are a more direct probe of galaxy properties as in we directly observe light from these galaxies as a function of their luminosity. Therefore, we will not be able to constrain properties of smaller, faint galaxies which are not observable (for example, $M_{\rm UV}\gtrsim -16$ in Fig. \ref{fig:fit_fiducial}).  The UVLF constraint becomes weaker compared to the $\tau$ constraint at high $k$ in Fig. \ref{fig:UVLF_constraint} due to this effect.

\section{Conclusions}
\label{Sec:conclusions}
In this work, we constrain modification to inflationary power spectrum at wavenumbers $0.3\lesssim k\lesssim 50$ Mpc$^{-1}$ using galaxy UVLF data from HST within redshift range of 4 to 8. The strongest constraint corresponds to $A_I\sim 10^{-9}$ for $k_{\rm peak}\sim 2$ Mpc$^{-1}$. Our constraints are complementary to previously obtained constraint using optical depth of reionization. However, UVLFs are a more direct probe as they directly count the number of galaxies as opposed to indirect effect of galaxies on the IGM.

The analysis presented here assumes a parameterized astrophysical model which may be not sufficient at higher redshifts. The preference for non-zero $A_I$ for $k_{\rm peak}$=1 ${\rm Mpc^{-1}}$ may have to do with insufficient astrophysical modelling which fails to capture all the relevant physical processes. Complementary cosmological observables such as 21cm global, power spectrum and intensity mapping experiments should also be sensitive to inflation power spectrum as well as star formation. A joint modelling of all these observables may further reduce the uncertainties in astrophysical modelling and tighten up the constraints or result in a detection.    

\section*{Acknowledgements}
We acknowledge discussions with Prof Pravabati Chingangbam during the course of this work.


{
\vspace{-3mm}
\bibliographystyle{unsrtads}
\bibliography{main}
}

\appendix


\begin{table}
\begin{subtable}[t]{0.48\textwidth}
\begin{tabular}[t]{l|r}
$M_{\rm UV}$ & $\Phi_{\rm UV}\hspace{0.1cm} {\rm (Mpc^{-3} mag^{-1})}$ \\
\hline
-22.69 & 0.000003$\pm$0.000004 \\
-22.19 & 0.000015$\pm$0.000009 \\
-21.69 & 0.000134$\pm$0.000023 \\
-21.19 & 0.000393$\pm$0.000040 \\
-20.69 & 0.000678$\pm$0.000063 \\
-20.19 & 0.001696$\pm$0.000113 \\
-19.69 & 0.002475$\pm$0.000185 \\
-19.19 & 0.002984$\pm$0.000255 \\
-18.69 & 0.005352$\pm$0.000446 \\
-18.19 & 0.006865$\pm$0.001043 \\
-17.69 & 0.010473$\pm$0.002229 \\
-16.94 & 0.024580$\pm$0.003500 \\
-15.94 & 0.025080$\pm$0.007860
\end{tabular}
\caption{z=4}
\label{tab:table1_a}
\end{subtable}
\hspace{\fill}
\begin{subtable}[t]{0.48\textwidth}
\begin{tabular}[t]{l|r}
$M_{\rm UV}$ & $\Phi_{\rm UV}\hspace{0.1cm} {\rm (Mpc^{-3} mag^{-1})}$ \\
\hline
-23.11 & 0.000002$\pm$0.000002 \\
-22.61 & 0.000006$\pm$0.000003 \\
-22.11 & 0.000034$\pm$0.000008 \\
-21.61 & 0.000101$\pm$0.000014 \\
-21.11 & 0.000265$\pm$0.000025 \\
-20.61 & 0.000676$\pm$0.000046 \\
-20.11 & 0.001029$\pm$0.000067 \\
-19.61 & 0.001329$\pm$0.000094 \\
-19.11 & 0.002085$\pm$0.000171 \\
-18.36 & 0.004460$\pm$0.000540 \\
-17.36 & 0.008600$\pm$0.001760 \\
-16.36 & 0.024400$\pm$0.007160

\end{tabular}
\caption{z=5}
\label{tab:table1_b}
\end{subtable}

\bigskip 

\begin{subtable}[t]{0.48\textwidth}
\begin{tabular}[t]{l|r}
$M_{\rm UV}$ & $\Phi_{\rm UV}\hspace{0.1cm} {\rm (Mpc^{-3} mag^{-1})}$ \\
\hline
-22.52 & 0.000002$\pm$0.000002 \\
-22.02 & 0.000015$\pm$0.000006 \\
-21.52 & 0.000053$\pm$0.000012 \\
-21.02 & 0.000176$\pm$0.000025 \\
-20.52 & 0.000320$\pm$0.000041 \\
-20.02 & 0.000698$\pm$0.000083 \\
-19.52 & 0.001246$\pm$0.000137 \\
-18.77 & 0.001900$\pm$0.000320 \\
-17.77 & 0.006680$\pm$0.001380 \\
-16.77 & 0.013640$\pm$0.004200

\end{tabular}
\caption{z=6}
\label{tab:table1_c}
\end{subtable}
\hspace{\fill}
\vspace{0.1cm}
\begin{subtable}[t]{0.48\textwidth}
\begin{tabular}[t]{l|r}
$M_{\rm UV}$ & $\Phi_{\rm UV}\hspace{0.1cm} {\rm (Mpc^{-3} mag^{-1})}$ \\
\hline
-22.16 & 0.000001$\pm$0.000002 \\
-21.66 & 0.000033$\pm$0.000009 \\
-21.16 & 0.000048$\pm$0.000015 \\
-20.66 & 0.000193$\pm$0.000034 \\
-20.16 & 0.000309$\pm$0.000061 \\
-19.66 & 0.000654$\pm$0.000100 \\
-19.16 & 0.000907$\pm$0.000177 \\
-18.66 & 0.001717$\pm$0.000478 \\
-17.91 & 0.005840$\pm$0.001460 \\
-16.91 & 0.008500$\pm$0.002940 

\end{tabular}
\caption{z=7}
\label{tab:table1_d}
\end{subtable}
\hspace{\fill}
\begin{subtable}[t]{0.48\textwidth}
\begin{tabular}[t]{l|r}
$M_{\rm UV}$ & $\Phi_{\rm UV}\hspace{0.1cm} {\rm (Mpc^{-3} mag^{-1})}$ \\
\hline
-21.87 & 0.000005$\pm$0.000003 \\
-21.37 & 0.000013$\pm$0.000005 \\
-20.87 & 0.000058$\pm$0.000015 \\
-20.37 & 0.000060$\pm$0.000026 \\
-19.87 & 0.000331$\pm$0.000104 \\
-19.37 & 0.000533$\pm$0.000226 \\
-18.62 & 0.001060$\pm$0.000340 \\
-17.62 & 0.002740$\pm$0.001040

\end{tabular}
\caption{z=8}
\label{tab:table1_e}
\end{subtable}
\caption{UVLF data between $z=4$ to 8 used in this work}
\label{tab:table1}
\end{table}

\begin{figure*}[]
    \begin{subfigure}[t]{1.0\textwidth}
        \centering
        \includegraphics[height=3.9in]{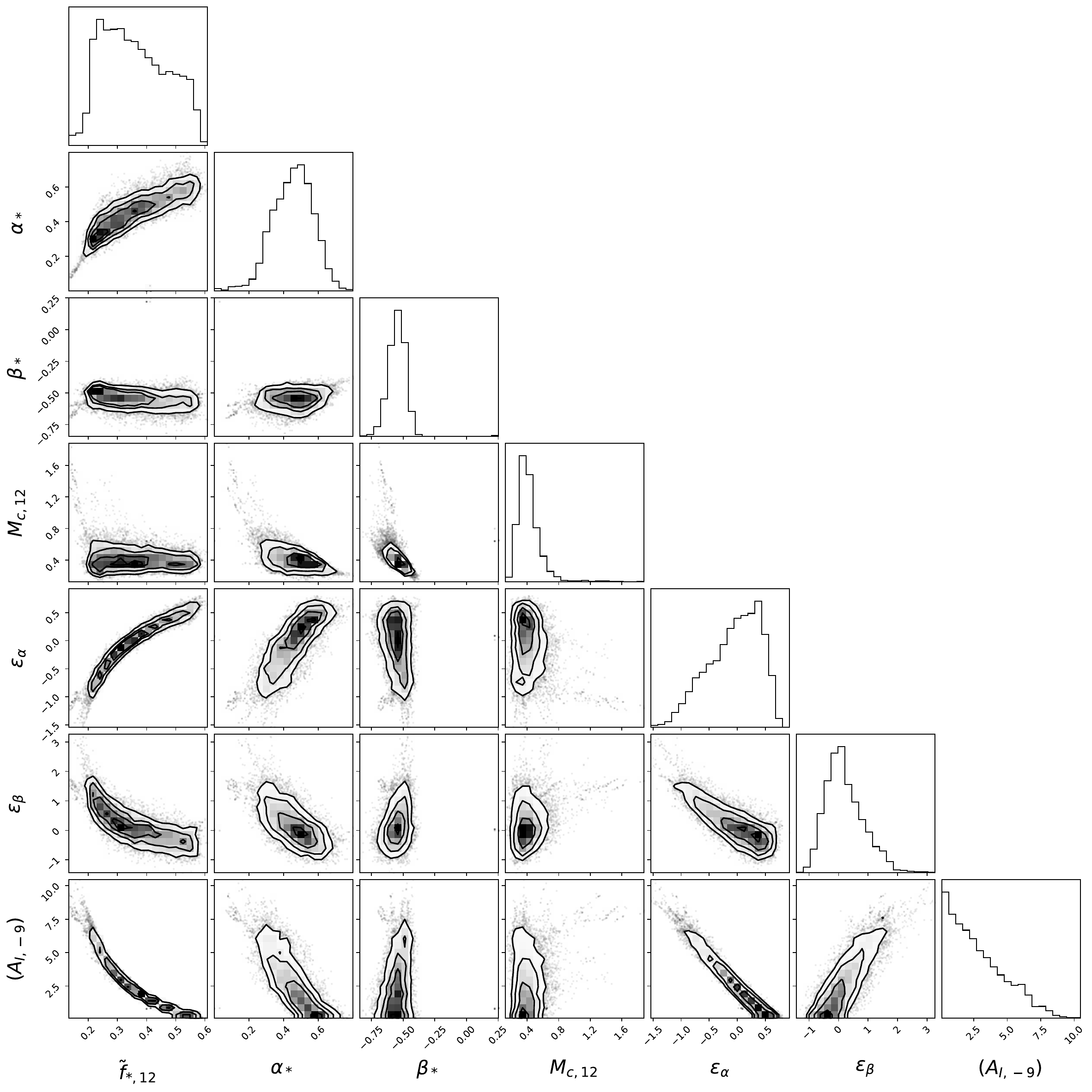}
    \end{subfigure}%
    \vspace{1cm}
    \begin{subfigure}[t]{1.0\textwidth}
        \centering
        \includegraphics[height=3.9in]{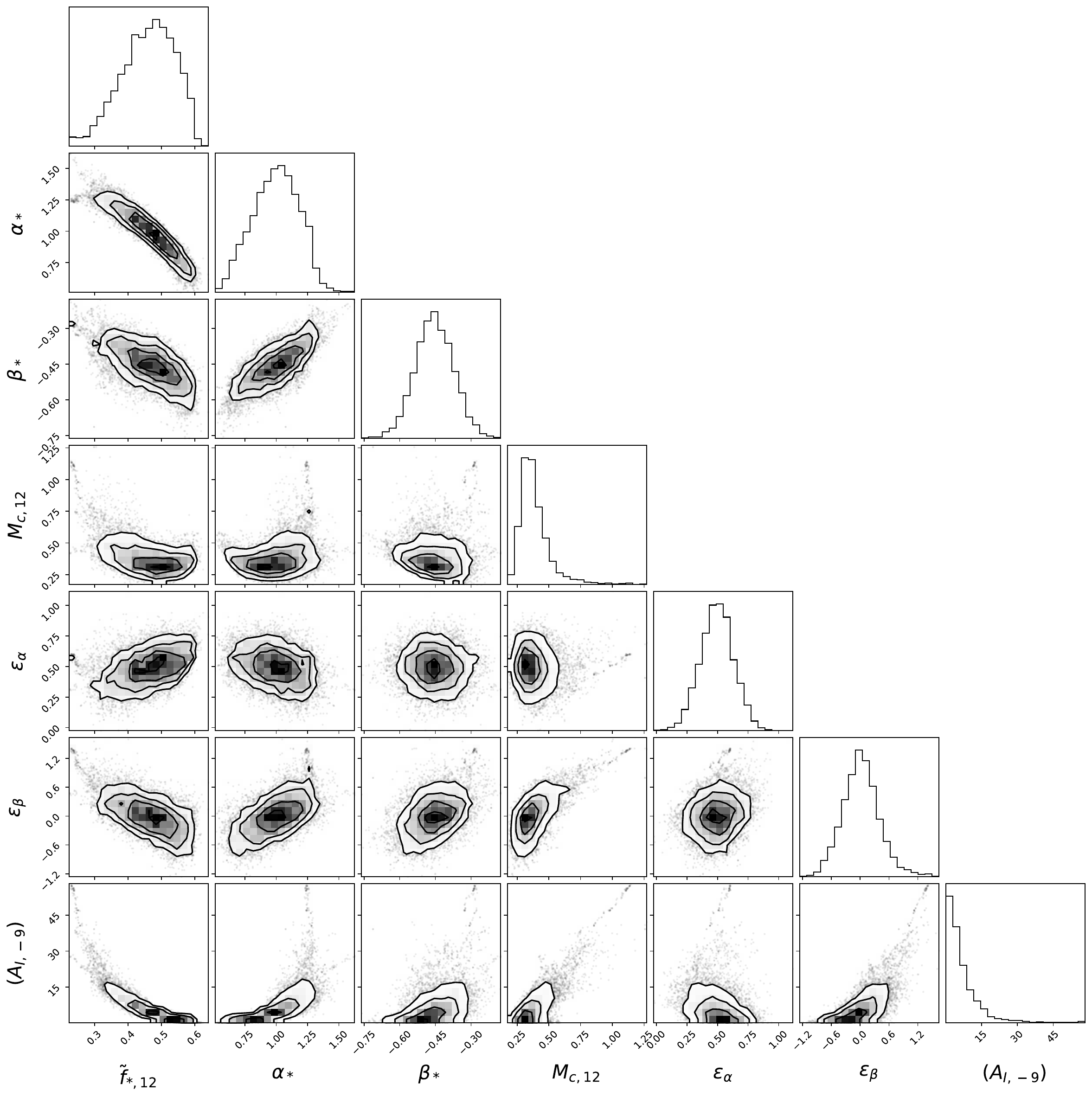}
    \end{subfigure}%
    \caption{Corner plot for our model with bump like feature at $k_{\rm peak}=0.5$ ${\rm Mpc^{-1}}$ (top) and $k_{\rm peak}=5$ ${\rm Mpc^{-1}}$ (bottom). }
    \label{fig:corner_k_0.5_10}
 \end{figure*}

\begin{figure*}[]
    \begin{subfigure}[t]{1.0\textwidth}
        \centering
        \includegraphics[height=3.9in]{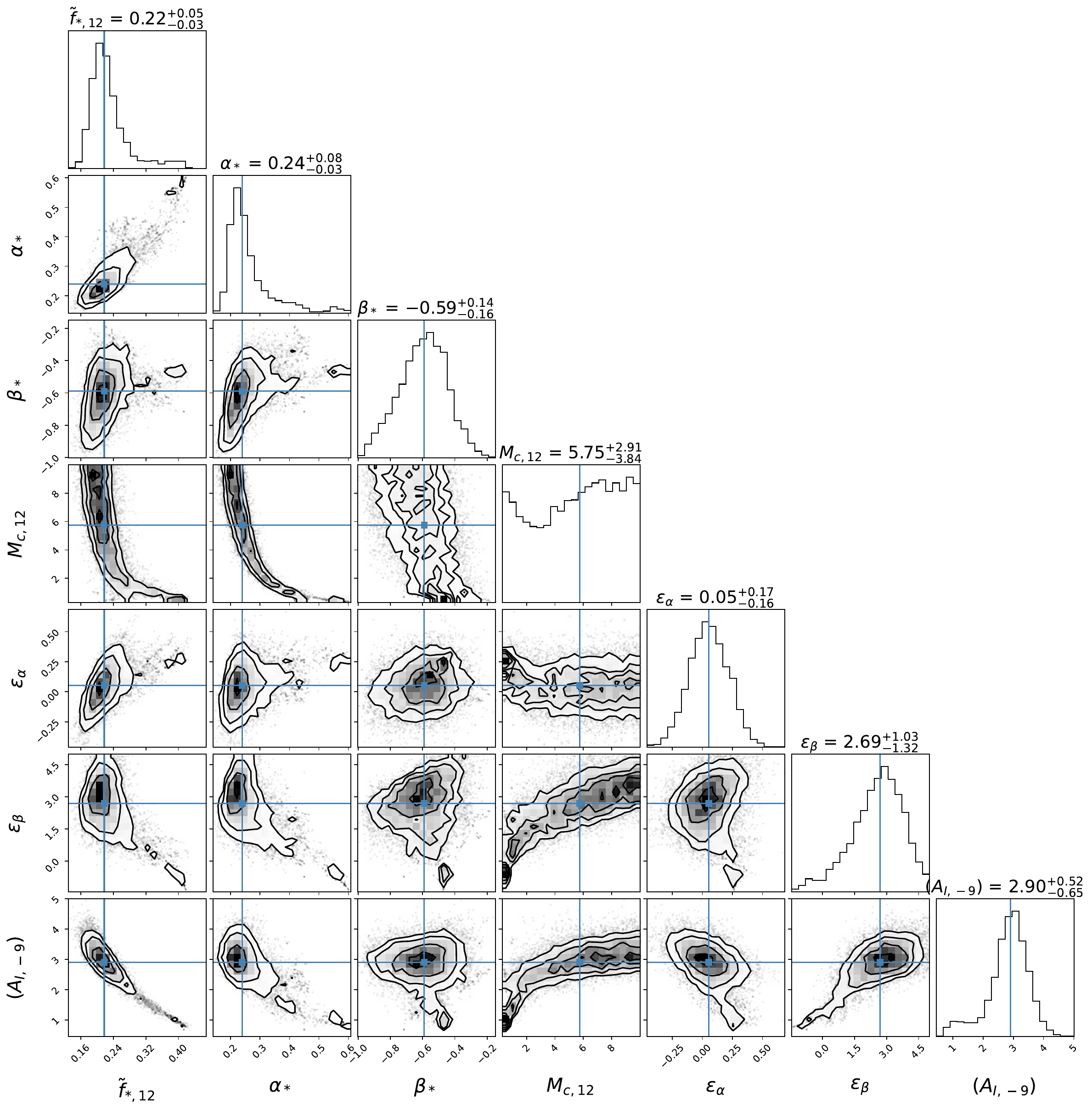}
    \end{subfigure}%
    \vspace{1cm}
    \begin{subfigure}[t]{1.0\textwidth}
        \centering
        \includegraphics[height=3.9in]{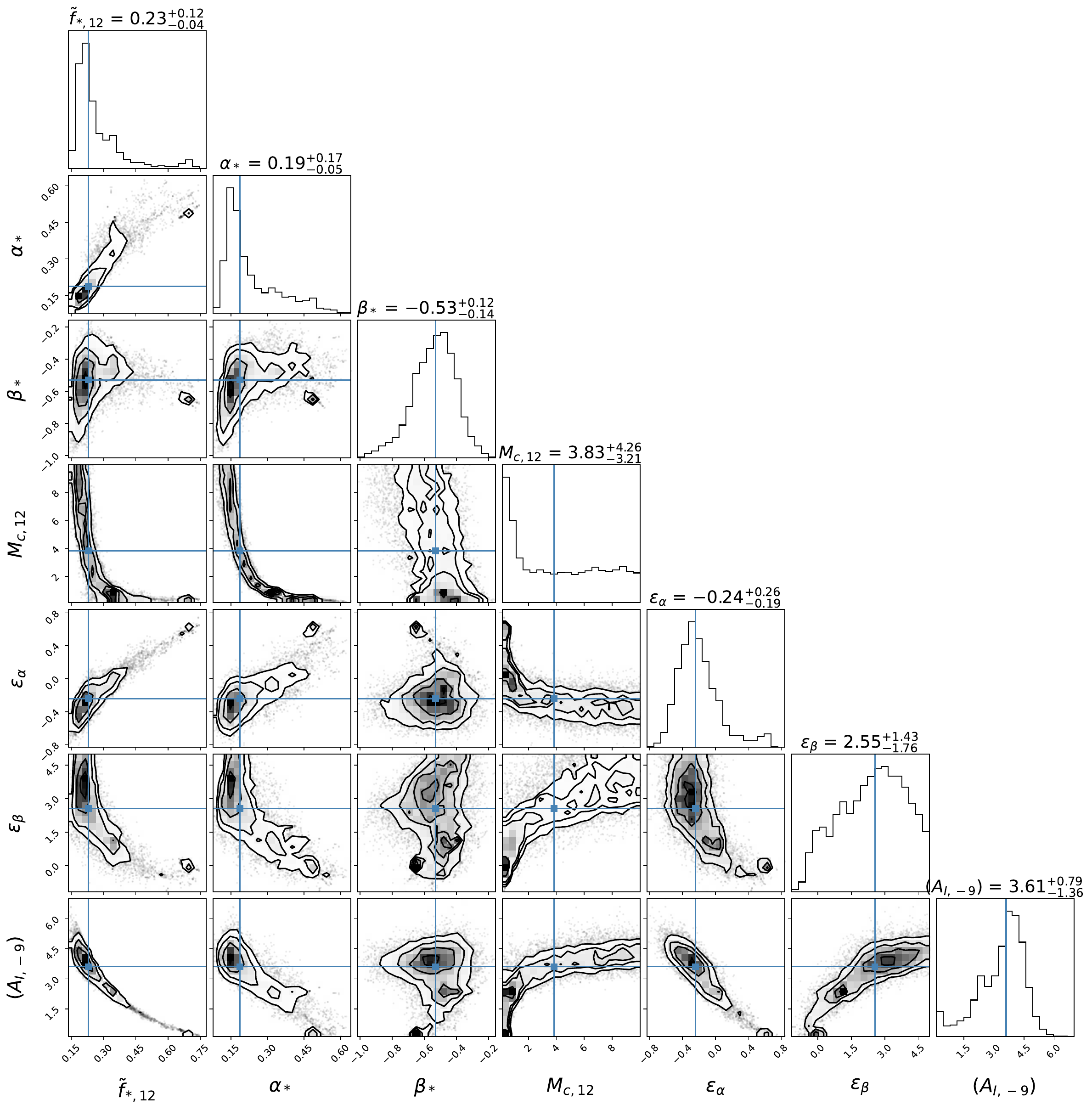}
    \end{subfigure}%
    \caption{Comparison of our results for $k_{\rm peak}=1$ ${\rm Mpc^{-1}}$ for Jenkins (top) and Tinker (bottom) halo mass function. We also denote the median and 2.7, 97.3 percentile region in the error bar to denote 95\% confidence region. }
    \label{fig:corner_k_1.0_comp}
 \end{figure*}

\end{document}